\documentclass[aps,prl,twocolumn,showpacs,preprintnumbers,amsmath,amssymb]{revtex4}   

\usepackage{graphicx}
\usepackage{dcolumn}
\usepackage{bm}

\allowdisplaybreaks

\newcommand {\beq} {\begin{equation}}
\newcommand {\eeq} {\end{equation}}
\newcommand {\beqa}{\begin{eqnarray}}
\newcommand {\eeqa}{\end{eqnarray}}

\newcommand {\ee}{\mbox{e}}

\newcommand{\idmat}{\mbox{\boldmath $1$}}

\begin{document}


\title{Deconfinement phase transition in
${\cal N}=4$ super Yang-Mills theory\\
on $R \times S^3$ from supersymmetric matrix quantum mechanics}
 
\author{Goro Ishiki$^{1,2}$}
\email{ishiki@post.kek.jp}
\author{Sang-Woo Kim$^{3}$}
\email{sangwookim@sogang.ac.kr}
\author{Jun Nishimura$^{2,4}$}
\email{jnishi@post.kek.jp}
\author{Asato Tsuchiya$^{5}$}
\email{satsuch@ipc.shizuoka.ac.jp}

\affiliation{
$^{1}$Department of Physics, 
Osaka University, Toyonaka, Osaka 560-0043, Japan\\
$^{2}$
High Energy Accelerator Research Organization (KEK), 
Tsukuba, Ibaraki 305-0801, Japan \\
$^{3}$Center for Quantum Spacetime (CQUeST), 
Sogang University, Seoul 121-742, Korea\\
$^{4}$Department of Particle and Nuclear Physics,
School of High Energy Accelerator Science,
Graduate University for Advanced Studies (SOKENDAI),
Tsukuba, Ibaraki 305-0801, Japan\\
$^{5}$Department of Physics, Shizuoka University,
836 Ohya, Suruga-ku, Shizuoka 422-8529, Japan
}

\date{October, 2008; preprint: OU-HET 613, KEK-TH-1280
}


\begin{abstract}
We test
the recent claim that
supersymmetric matrix quantum mechanics
with mass deformation preserving maximal supersymmetry
can be used to study
${\cal N}=4$ super Yang-Mills theory
on $R \times S^3$ 
in the planar limit.
When the mass parameter is large, we can integrate
out
all the massive fluctuations 
around a particular classical solution,
which corresponds to $R \times S^3$.
The resulting effective theory
for 
the gauge field moduli
at finite temperature
is studied both analytically and numerically,
and shown to reproduce the deconfinement phase transition in
${\cal N}=4$ 
super Yang-Mills theory 
on $R \times S^3$ at weak coupling.
This transition was 
speculated to be
a continuation of the conjectured phase transition at strong coupling,
which corresponds to the Hawking-Page transition
based on the gauge-gravity duality.
%
%
%
By choosing a different classical solution of the same model,
one can also reproduce results for 
gauge theories on other space-time such as
$R \times S^3/Z_q$ and $R \times S^2$.
All these theories can be studied at strong coupling
by the new simulation method, which was used successfully
for supersymmetric matrix quantum mechanics without mass deformation.
%
\end{abstract}




\maketitle


\paragraph*{Introduction.---}

Supersymmetric large-$N$ gauge theories have attracted
much attention in the past decade
due to its conjectured duality \cite{Maldacena:1997re}
to classical gravity theories.
In order to 
investigate the duality further and to make use of it,
one has to study the gauge theories in the
strongly coupled regime.
Monte Carlo simulation can be a powerful approach
as in QCD, but the
problem is that
the naive lattice regularization breaks supersymmetry.
Recently there have been considerable developments 
in constructing lattice theories preserving \emph{part of}
supersymmetry \cite{latticeSUSY}.
%
However, in some cases including theories
of interest in the context of gauge-gravity duality,
%
there might be 
a regularization scheme alternative to the lattice,
which respects supersymmetry maximally.
In the case of 1d gauge theory
or matrix quantum mechanics (MQM), for instance,
one can fix the gauge nonperturbatively and introduce
a Fourier mode cutoff \cite{Hanada:2007ti}.
Using this method,
the case with 16 supercharges has been simulated
at various coupling constant \cite{AHNT},
and the results interpolated nicely 
the weak-coupling behavior obtained 
by high temperature expansion \cite{Kawahara:2007ib}
and the strong-coupling behavior predicted
from the gauge-gravity duality \cite{Itzhaki:1998dd}.
(Consistent results
were obtained \cite{Catterall:2008yz} also by 
using the lattice approach  \cite{Catterall:2007fp}.)
The aim of this Letter is to make a first step
towards an extension of this work to higher dimensions.

In Ref.\ \cite{Ishii:2008ib}
the use of
supersymmetric 
MQM
with mass deformation preserving supersymmetry
has been proposed as a regularization
of 
${\cal N}=4$ 
super Yang-Mills theory (SYM)
on $R \times S^3$ 
in the planar limit.
In fact the supersymmetry of the target theory
is enhanced to the superconformal symmetry,
which includes 32 supercharges.
%
The regularized theory has only 16 supercharges, but
we consider it optimal since the conformal
symmetry is necessarily broken by any regularizations.
We test this proposal
when the mass parameter is large, in which case
we can integrate out
all the massive fluctuations 
around a particular classical solution,
which corresponds to $R \times S^3$.
The resulting effective theory
for the gauge field moduli
at finite temperature
%
reproduces the deconfinement phase transition in
${\cal N}=4$ 
SYM
on $R \times S^3$ at weak coupling.
This transition \cite{Sundborg:1999ue,Aharony:2003sx}
was speculated
to be a continuation of the conjectured
phase transition at strong coupling \cite{Witten:1998zw}, which
corresponds to the Hawking-Page transition \cite{Hawking:1982dh}
based on the gauge-gravity duality.

\paragraph*{The model and the classical vacuum.---}
The model we study is defined by \cite{Berenstein:2002jq}
\begin{align}
S
= &\frac{1}{g
^2}
\int
dt \, \mbox{tr} 
\left[\frac{1}{2}(D_tX_M)^2-\frac{1}{4}[X_M,X_N]^2 \right.\nonumber\\
&
+\frac{1}{2}\Psi^{\dagger} D_t \Psi
-\frac{1}{2}\Psi^{\dagger}\gamma_M[X_M,\Psi]+
\frac{\mu^2}{2}(X_i)^2 \nonumber\\
&\left. +\frac{\mu^2}{8}(X_a)^2 +i\mu\epsilon_{ijk}X_iX_jX_k
+i\frac{3\mu}{8}\Psi^{\dagger}\gamma_{123}\Psi \right],
\label{pp-action}
\end{align}
where the range of indices is given by
$1 \le M,N \le 9$, $1 \le i,j,k \le 3$ and $4 \le a \le 9$.
The covariant derivative is defined by
$D_t=\partial_t-i[A, \cdot]$,
where $A$ as well as $X_M$ and $\Psi$ in (\ref{pp-action})
are $N\times N$ matrices
depending on $t$. 
The model has SU$(2|4)$ symmetry, which includes
16 supercharges, and 
for $\mu=0$ it reduces to the supersymmetric 
MQM
studied recently by Monte Carlo 
simulation \cite{AHNT,Catterall:2008yz}.

The model possesses many vacua
\begin{align}
X_i=\mu \bigoplus_{I=1}^{\nu}(L_i^{(n_I)}\otimes \idmat_{k_I} ) 
\label{background}
\end{align}
representing multi fuzzy spheres, 
where $L_i^{(n_I)}$ are 
the $n_I$-dimensional irreducible representation of
the SU$(2)$ generators obeying
$[L_i^{(n_I)},L_j^{(n_I)}]=i\epsilon_{ijk}L_k^{(n_I)}$. 
The parameters $n_I$ and $k_I$ in (\ref{background}) satisfy
the relation $\sum_{I=1}^{\nu}n_Ik_I=N$.
These vacua preserve the SU$(2|4)$ symmetry and they are all 
degenerate.
Let us consider
the theory (\ref{pp-action})
around the background (\ref{background})
in the case
\begin{align}
k_I=k \ , \quad n_I=n+I-\frac{\nu+1}{2} 
\label{our background}
\end{align}
for $I =1, \cdots , \nu$, assuming $\nu$ to be odd,
and take the limit
\begin{align}
&n \rightarrow\infty, \;\; \nu\rightarrow\infty, \;\;k\rightarrow \infty,\;\;
n-\frac{\nu}{2}\rightarrow\infty\nonumber\\
&\mbox{with} \;\; 
\frac{g^2 k}{n}=\frac{\lambda}{v}=\mbox{fixed} \ ,
\label{limit}
\end{align}
where 
$v= 2\pi^2 (2/\mu)^3$.
It is claimed \cite{Ishii:2008ib} 
(See also Refs.\ \cite{Ishiki:2006yr,Ishii:2008tm} for earlier
proposals.) that the resulting theory is actually equivalent
to the planar limit of 
${\cal N}=4$ SYM on $R \times S^3$,
where $\lambda$ is
the 't Hooft coupling constant
and the radius of $S^3$ is given by $2/\mu$.
The parameter $v$ introduced
above
is just 
the volume of the $S^3$.

\paragraph*{The argument for the equivalence.---}

Let us briefly review the argument for the equivalence \cite{Ishii:2008ib}.
Here $S^3$ is regarded as an $S^1$ bundle over $S^2$. 
By making the Kaluza-Klein (KK) reduction 
along the $S^1$ fiber direction,
one obtains the KK modes on $S^2$.
Reflecting the non-trivial fiber structure, however,
the KK modes
should be expanded in terms of
monopole harmonics on $S^2$ ---which are not single-valued in
general unlike ordinary spherical harmonics---
in the presence of the monopole charges
corresponding to the KK momenta.
On the other hand, the ``fuzzy regularization'' of the 
monopole harmonics is given by $(n+2q) \times n$ 
rectangular matrices, where $q$ represents the monopole charge.
The fluctuations in the $(I,J)$ block 
around the background (\ref{background})
for the case (\ref{our background})
can therefore be regarded as the KK mode
with the KK momentum $(I-J)/2$. 
Note that we obtain precisely the right KK momentum spectrum
albeit with the cutoff $(\nu -1)/2$.
%
%
The parameter $n$ plays the role of the cutoff
for the angular momentum on $S^2$.
Let us emphasize that
these two momentum cutoffs preserve
the gauge invariance 
and the SU$(2|4)$ symmetry, which is a subgroup
of the SU$(2,2|4)$ superconformal symmetry.

%

Except for the existence of the cutoffs,
planar diagrams obtained by expanding
the supersymmetric 
MQM
around the background (\ref{our background})
agree with planar diagrams in the
${\cal N}=4$ SYM on $R \times S^3$.
Nonplanar diagrams do not agree for two reasons.
One is that the $S^2$ is constructed as a fuzzy sphere,
and the fuzziness affects nonplanar diagrams. 
The other is that the reduction in the $S^1$ direction
is expected to occur due to the mechanism analogous to the 
quenched Eguchi-Kawai model \cite{Eguchi:1982nm,BHN},
which works only for planar diagrams.
The role of the quenched momentum variables 
are played
by the monopole charges.
Unlike in the usual quenched Eguchi-Kawai model \cite{BHN},
however, the momentum has a discrete spectrum
corresponding to $S^1$.
Due to this difference, sending $\nu$ to infinity alone does not
remove nonplanar diagrams,
and therefore one needs to take the large-$k$ limit.

Let us also emphasize that
the model (\ref{pp-action}) is a massive theory,
which has no flat direction.
Furthermore,
the background (\ref{our background}) is stable against
quantum fluctuations thanks to the SU$(2|4)$ symmetry 
at least at zero temperature.
Tunneling to the other vacua through the instanton 
effects is suppressed in the large-$k$ limit.
Hence there is no need to do something like 
momentum quenching, which is necessary 
in the quenched Eguchi-Kawai model.

%
%

\paragraph*{Effective theory for the gauge field moduli.---}
Let us introduce finite temperature $T$ by 
compactifying the Euclidean time $t$ in (\ref{pp-action})
to a circle with the circumference $\beta=T^{-1}$.
Unlike the $T=0$ case, one cannot set the gauge field to zero 
due to the nontrivial holonomy along the $t$ direction.
In fact the gauge field contains 
some moduli given by
$A(t)=\bigoplus_{I=1}^{\nu} 
\{ \idmat_{n_I}\otimes \bar{A}^{(I)}(t) \} $
%
around the general background (\ref{background}), 
where $\bar{A}^{(I)}(t)$ are $k_I \times k_I$ hermitian matrices.
One can choose a gauge 
in which $\bar{A}^{(I)}(t)$ takes the form
$\bar{A}^{(I)}(t)=
\frac{1}{\beta} \, 
\mbox{diag} ( \alpha^{(I)}_1,\cdots,\alpha^{(I)}_{k_I}
) $,
where $\alpha^{(I)}_{a} \in (-\pi , \pi]$ ($a=1, \cdots , k_I$).
%
At small $g^2$ or 
at large $\mu$,
one can integrate over all the massive fluctuations
around the general background (\ref{background}),
and the effective action for the gauge field moduli
is given by eq.\ (3.10) in 
Ref.\ \cite{Kawahara:2006hs}.

Here we restrict ourselves to 
the particular case (\ref{our background}),
and study the effective action
analytically in the large-$k$ limit
to test the equivalence to the planar limit of
${\cal N}=4$ SYM on $R \times S^3$
at small $\lambda$ (See eq.\ (\ref{limit}).).
%
%
For that purpose we rewrite the effective action
in terms of the distribution functions of $\alpha_a^{(I)}$ defined by
\begin{align}
\rho^{(I)}(\theta)=\frac{1}{k}\sum_{a=1}^k\delta(\theta-\alpha_a^{(I)}) \ .
\end{align}
One can easily arrive at the form
%
\begin{align}
 S_{\rm MQM}  =&  k^2 \!\! \sum_{I,J=1}^{\nu}
\int
d\theta d\theta' 
\rho^{(I)}(\theta)V^{(I,J)}(\theta-\theta')\rho^{(J)}(\theta') \ ,  
\label{effective action for matrix model} \\
 V^{(I,J)}(\theta) =&
\sum_{p=1}^{\infty} \tilde{V}^{(I,J)}_p\cos(p\theta) \ ,  \\
 \tilde{V}^{(I,J)}_p =&
\frac{1}{p}
\Bigl\{
\delta_{IJ}-6z^{(I,J)}_{s}(x^p)-z^{(I,J)}_{v}(x^p)  \nonumber \\
&  -4(-1)^{p+1}z^{(I,J)}_{f}(x^p) \Bigr\} \ ,
\label{Vtilde}
\end{align}
where we have introduced a dimensionless parameter
\beq
x = \ee ^{ - \beta \mu /2 }
\label{defx}
\eeq
and the functions
\begin{align}
z_{s}^{(I,J)}(x)
=&x\frac{\partial}{\partial x}\left(
\frac{x^{|n_I-n_J|+1}(1-x^{n_{IJ}})}{1-x^2}\right),
\label{zsp} \\
z_{v}^{(I,J)}(x)=&x^2\frac{\partial}{\partial x}
\left(\frac{x^{|n_I-n_J|-1+2\delta_{IJ}}(1-x^{n_{IJ}-2\delta_{IJ}})}
{1-x^2}\right) \nonumber\\
&+\frac{\partial}{\partial x}
\left(\frac{x^{|n_I-n_J|+3}(1-x^{n_{IJ}})}
{1-x^2}\right),
\label{zvp} \\
z_{f}^{(I,J)}(x)=&x^{\frac{3}{2}}\frac{\partial}{\partial x}
\left(\frac{x^{|n_I-n_J|+2\delta_{IJ}}(1-x^{n_{IJ}-2\delta_{IJ}})}
{1-x^2}\right) \nonumber\\
&+x^{\frac{1}{2}}\frac{\partial}{\partial x}
\left(\frac{x^{|n_I-n_J|+2}(1-x^{n_{IJ}})}
{1-x^2}\right) 
\label{zfp} 
\end{align}
with $n_{IJ}=n_I+n_J-|n_I-n_J|$,
which can be interpreted as
the single-particle partition functions 
for the scalars, the vector and the fermions, respectively.

Since $k$ appears only as an overall coefficient in the
effective action (\ref{effective action for matrix model}),
one can solve the theory exactly in the large-$k$ limit 
by the saddle-point equation
\begin{align}
\sum_{J=1}^{\nu}
\int
d\theta' \, {V^{'(I,J)}}
(\theta-\theta') \, \rho^{(J)}(\theta')  
=0 
\ .
\label{saddle-point equation for matrix model}
\end{align}
The free energy is obtained by
$F_{\rm MQM}  = T \, S_{\rm MQM}$,
where we use the solution which minimizes the action
when we evaluate $S_{\rm MQM}$ on the right hand side.

\paragraph*{Results for ${\cal N}=4$ SYM on $R \times S^3$.---}


To proceed further, let us recall some known results
for ${\cal N}=4$ U($k$) SYM on $R \times S^3$
in the planar large-$k$ limit at weak coupling.
Integrating out all the massive fields,
one obtains the effective action
\cite{Sundborg:1999ue,Aharony:2003sx}
\begin{align}
S_{\rm SYM}=k^2\int d\theta d\theta' 
\rho(\theta)V(\theta-\theta')\rho(\theta')  
\label{continuum effective action}
\end{align}
for the distribution $\rho(\theta)$
of the gauge field moduli, 
where the kernel $V(\theta)$ is expanded as
\begin{align}
\label{Vdef}
&V(\theta) =
\sum_{p=1}^{\infty}\tilde{V}_p\cos(p\theta) \  , \\
&\tilde{V}_p =
\frac{1}{p}\Bigl\{1-6z_s(x^p)-z_v(x^p)
-4(-1)^{p+1}z_f(x^p) \Bigr\} \ .
\nonumber
\end{align}
The single partition functions are written as
\beq
z_s(x)=\frac{x+x^2}{(1-x)^3}, 
z_v(x)=\frac{6x^2-2x^3}{(1-x)^3} , 
z_f(x)= \frac{4x^{\frac{3}{2}}}{(1-x)^3} \ . 
\nonumber
\eeq


Obviously the uniform distribution is always a solution
to 
the saddle-point equation.
At low temperature, it gives the absolute minimum of
the effective action.
One can show that there is a first order phase transition
at a critical point determined 
by $\tilde{V}_1=0$ 
as $x_c=7-4\sqrt{3}$ \cite{Sundborg:1999ue,Aharony:2003sx}
in terms of the dimensionless parameter (\ref{defx}).
%
%
%
Above the critical temperature, 
the dominant solution has a compact support 
$[-\theta_0,\theta_0]$ with $\theta_0<\pi$,
and near the critical temperature, in particular, 
its explicit form is given by the Gross-Witten form
\cite{Sundborg:1999ue,Aharony:2003sx}
\begin{align}
& \rho (\theta)=\frac{1}{\pi \omega} 
\left(\cos \frac{\theta}{2}\right)
\sqrt{ \omega
-\sin^2 \frac{\theta}{2} }  
\label{eigen_anal}
\end{align}
for $|\theta| \le \theta_0$, where 
$\theta_0=2 \sin ^{-1} \sqrt{\omega}$ and
$\omega = 
1-\sqrt{\frac{-\tilde{V}_1}{1-\tilde{V}_1}} $.


\paragraph*{Results for MQM.---}

First let us note that the kernel $V^{(I,J)}(\theta)$ in
(\ref{effective action for matrix model})
decreases exponentially as $|I-J|$ becomes large.
Moreover, one finds that
\begin{align}
\sum_{J=1}^{\nu}V^{(I,J)}(\theta)=V(\theta)+\Delta V^{(I)}(\theta) \ ,
\label{sum over J}
\end{align}
where $V(\theta)$ is the kernel (\ref{Vdef})
for the ${\cal N}=4$ SYM,
and the remaining $I$-dependent part
$\Delta V^{(I)}(\theta)$ 
decreases exponentially as one moves away from
the edges $I=1$ and $I=\nu$.
This is the case since
$\sum_{J=1}^{\nu} z_{i}^{(I,J)}(x) = z_i(x) + \cdots $
for $i=s,v,f$, where the dots represent terms
proportional to positive powers of $x^I$ and $x^{\nu-I+1}$.
Thus we may naturally expect 
the solution 
to (\ref{saddle-point equation for matrix model}) 
to be
\begin{align}
\rho^{(I)}(\theta)=\hat{\rho}(\theta)+\Delta\rho^{(I)}(\theta) \ ,
\label{solution to saddle-point equation}
\end{align}
where $\hat{\rho}(\theta)$ is the solution
for the ${\cal N}=4$ SYM,
and $\Delta\rho^{(I)}(\theta)$ decreases exponentially
as one moves away from $I=1$ and $I=\nu$.
(See below for an explicit confirmation by 
simulation.)
Substituting (\ref{solution to saddle-point equation})
into (\ref{effective action for matrix model}),
we obtain
$\frac{1}{\nu k^2} S_{\rm MQM}
= \int
d\theta d\theta' 
\hat{\rho}(\theta)V(\theta-\theta')\hat{\rho}(\theta') + \cdots $,
where the abbreviated terms vanish as $1/\nu$
in the limit (\ref{limit}).
Thus in that limit we obtain the relationship
\begin{align}
\frac{1}{k^2 \nu} F_{\rm MQM} = \frac{1}{k^2} F_{\rm SYM} \ .
\label{agreement of free energy}
\end{align}
In fact the condition
$n-\nu/2\rightarrow\infty$ in (\ref{limit})
is not needed in deriving (\ref{agreement of free energy}).
Therefore it may not be necessary for the equivalence.

We can also show \cite{IKNT} that the critical temperatures
of the two theories agree 
without assuming the 
property (\ref{solution to saddle-point equation}).
The critical temperature
of our effective theory (\ref{effective action for matrix model})
can be determined by
$\det \tilde{V}^{(I,J)}_p=0$
for some $p$.
By noticing that $\tilde{V}^{(I,J)}_p$ becomes a Toeplitz matrix
in the $n\rightarrow \infty$ limit,
and using a property of its minimum eigenvalue
in the $\nu\rightarrow \infty$  limit \cite{Gray},
we obtain precisely the same condition 
$\tilde{V}_1=0$ for the critical point
of the effective theory (\ref{continuum effective action})
for the ${\cal N}=4$ SYM.
%

\paragraph*{Monte Carlo Simulation.---}

\begin{figure}[htb]
\begin{center}
\includegraphics[height=5.5cm]{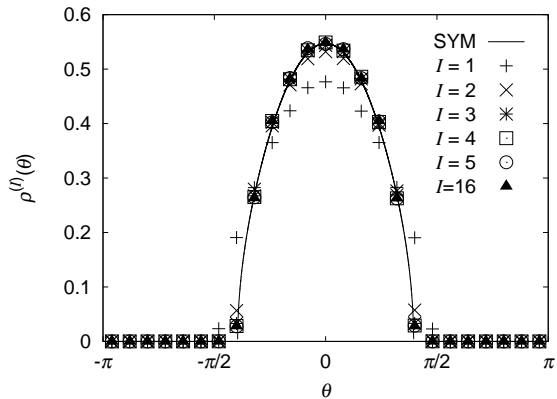}
\end{center}
\caption{
The distribution $\rho^{(I)}(\theta)$
of the gauge field moduli in the supersymmetric
MQM around the background (\ref{our background})
is plotted for $I=1,2,3,4,5,16$
with $k=16$ and $n=\nu=31$
at temperature corresponding to $x=0.104$ 
near the critical point $x_c=0.072$.
The statistical errors 
are smaller than the symbol size.
The solid line represents the result (\ref{eigen_anal})
for the ${\cal N}=4$ SYM on $R \times S^3$
at the same temperature.
}
\label{eigenval_distr}
\end{figure}

As is done in Ref.\ \cite{Kawahara:2006hs} for simpler cases,
we can perform Monte Carlo simulation of the effective theory
for the gauge field moduli in our case (\ref{our background}).
For instance, we have confirmed the statement
(\ref{agreement of free energy}) explicitly \cite{IKNT}.
%
%
Also we have checked that the distribution
$\rho^{(I)}(\theta)$ of the gauge field moduli
has the expected property
(\ref{solution to saddle-point equation}).
Figure \ref{eigenval_distr} shows 
%
%
that as one goes towards the midpoint $I=(\nu+1)/2$,
the distribution converges rapidly 
to the result (\ref{eigen_anal})
for the ${\cal N}=4$ SYM on $R \times S^3$
represented by the solid line.

\paragraph*{Summary and outlook.---}
In this Letter we presented explicit computations
at weak coupling and at finite temperature
that confirm the equivalence between the
supersymmetric 
MQM 
around a particular background
and the ${\cal N}=4$ SYM on $R \times S^3$
in the planar limit \cite{endnote}.
We may therefore hope to study 
this 4d theory
at various 't Hooft coupling constant
by simulating supersymmetric MQM
as has been done already for $\mu=0$ \cite{AHNT,Catterall:2008yz}.

By expanding 
the same model around
a different classical solution,
we also reproduce \cite{IKNT} analogous phase transitions in
supersymmetric gauge theories on other space-time such as
$R \times S^3/Z_q$ \cite{Hikida:2006qb}
and $R \times S^2$ \cite{Grignani:2007xz}.
We can study these theories also
at strong coupling and compare the results with
their gravity duals \cite{Lin:2005nh}.

The idea that dimensionally reduced large-$N$ gauge theories
can retain information of the theory before reduction dates back
to early eighties \cite{Eguchi:1982nm}.
Recently it is commonly considered in the opposite way
that actually 
the reduced models
are more fundamental,
and that the space \cite{BFSS} and possibly also time \cite{IKKT}
are merely an emergent notion.
Our results show that
the reduced model
reproduces gauge theory results 
in various space-time depending on which vacuum
one expands the theory around.
We consider that this sheds light on the aspects
of reduced models as a background independent
formulation of quantum gravity.
%



\paragraph*{Acknowledgments.---}
The work of G.\ I.\ 
is supported 
by JSPS.
The work of S.-W.\ K.\ 
is supported by the Center for Quantum Spacetime of Sogang University
(Grant number: R11-2005-021).
The work of J.\ N.\ and A.\ T.\ 
is supported 
by Grant-in-Aid for Scientific
Research (Nos.\ 19340066, 20540286 and 19540294) 
from the Ministry 
of Education, Culture, Sports, Science and Technology.





\begin{thebibliography}{99}
\bibitem{endnote}
When we were about to complete this work, 
we were informed that
the equivalence has been confirmed also
in the high $T$ limit
including higher loop corrections \cite{Kitazawa:2008mx}.
\end{thebibliography}
\end{document}